\documentclass[prb, twocolumn, floatfix]{revtex4}

\usepackage{graphicx}
\usepackage{epstopdf}

\begin{document}

\title{Approaching ideal weak link behavior with three dimensional aluminum nanobridges}
\author{R. Vijay}
\author{E. M. Levenson-Falk}
\author{D. H. Slichter}
\author{I. Siddiqi}
\email[Corresponding author: ]{irfan@berkeley.edu}
\affiliation{Quantum Nanoelectronics Laboratory, Department of Physics, University of California, Berkeley CA 94720}

\date{\today}

\begin{abstract}

We present transport measurements of unshunted dc superconducting quantum interference devices (SQUIDs) consisting of 30 nm wide aluminum nanobridges of varying length L contacted with two and three dimensional banks. 3D nanobridge SQUIDs with L $\leq$ 150 nm (approximately 3-4 times the superconducting coherence length) exhibit $\approx 70 \%$ critical current modulation with applied magnetic field, approaching the theoretical limit for an ideal short metallic weak link.  In contrast, 2D nanobridge SQUIDs exhibit significantly lower critical current modulation. This enhanced nonlinearity makes 3D nanobridge Josephson junctions well suited to optimize sensitivity in weak link SQUID magnetometers as well as realize ultra low-noise amplifiers and qubits. 
\end{abstract}


\maketitle

Josephson junction circuits based on weak links are well suited for applications requiring high transparency, small area junctions. For example, superconducting quantum interference devices (SQUIDs) constructed from nanoscale bridges are widely used in magnetization measurements of magnetic crystals and ferromagnetic clusters \cite{CSIRO, Hilgenkamp, Wernsdorfer_review}. Furthermore, weak link junctions are attractive candidates for coherent quantum circuits since, by construction, they are free of lossy tunnel oxide\textemdash a potential source of decoherence \cite{Martinis}. However, these performance gains come at the expense of reduced nonlinearity in the current-phase relation (CPR) when compared to conventional tunnel junctions. The CPR in weak link devices depends strongly on the junction geometry \cite{Likharev}.  In particular, thin film bridges with lateral dimensions comparable to the superconducting coherence length $\xi$ are predicted to have maximal nonlinearity when contacted with banks of greater thickness than the bridge (3D), as opposed to banks of equal thickness (2D).  The 3D banks enhance nonlinearity by acting as good phase reservoirs and confining the variation of the superconducting order parameter to the weak link \cite{vijay_wire}. 

In this letter, we present measurements of the magnetic flux response of unshunted dc SQUIDs comprised of aluminum nanobridges connected to 2D and 3D banks. Our measurements show enhanced critical current modulation in 3D nanoSQUIDs and confirm that thin film aluminum nanobridges up to 150 nm in length ($\approx 3 - 4 \xi$) with thick contacts approach the nonlinear CPR of an ideal, short, metallic weak link. Longer devices and those with 2D banks exhibit reduced modulation, in quantitative agreement with our numerically computed CPRs. Thus, 3D nanobridge junctions are robust nonlinear elements for use in nanoSQUID magnetometers, amplifiers, and coherent circuits such as quantum bits. 

\begin{figure}[htbp]
\includegraphics{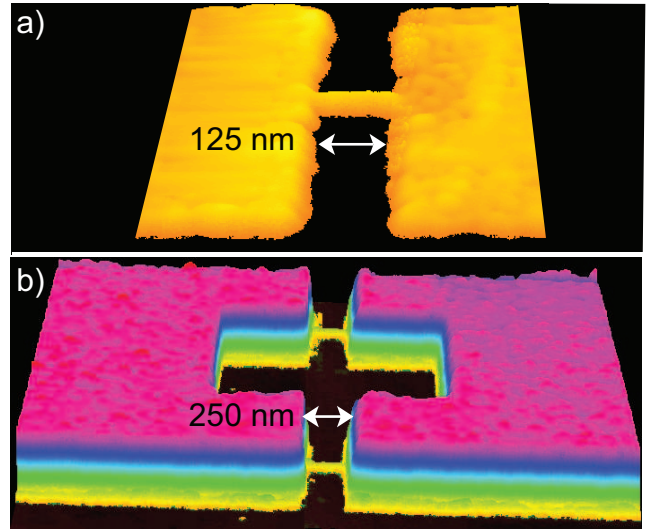}
\caption{\label{fig1} (Color online) Atomic force micrographs of 8 nm thick, 30 nm wide aluminum nanobridges contacted with (a) 8 nm thick 2D banks and (b) 80 nm thick 3D banks. Two nanobridge junctions are incorporated into an unshunted dc SQUID, as shown in (b).  The SQUID loop is 1.0 x 1.5 $\mu$m, and is calculated to have 5 pH of inductance.}
\end{figure}

Our devices consist of a small superconducting loop interrupted by two nanobridges. Each bridge is an 8 nm thick, 30 nm wide aluminum film of variable length from 75-400 nm. These dimensions can be reliably achieved using electron beam lithography and a liftoff process. For 2D devices, a single electron-beam evaporation is performed at normal incidence through a bilayer shadow mask to simultaneously define a narrow bridge and wide banks of the same thickness. To create 3D banks, we tilt the sample \textit{in situ} and perform a second evaporation step to deposit 80 nm thick banks without depositing further material in the narrow bridge region. Atomic force micrographs of 2D and 3D nanobridges are shown in Fig. 1. The SQUID loop is 1.0 x 1.5 $\mu$m, and is calculated to have $L_{loop}=5$ pH of geometric inductance. Each chip contained six simultaneously fabricated nanoSQUIDs with bridges of varying lengths. The devices were mounted in a copper sample box equipped with a superconducting flux-modulation coil and measured in a $^{3}$He cryostat with a base temperature of 265 mK. All wiring ran though discrete-element and distributed lossy filters to suppress spurious electromagnetic radiation. Four-wire transport measurements on all six SQUIDs were performed using custom electronics.

\begin{figure}
\includegraphics{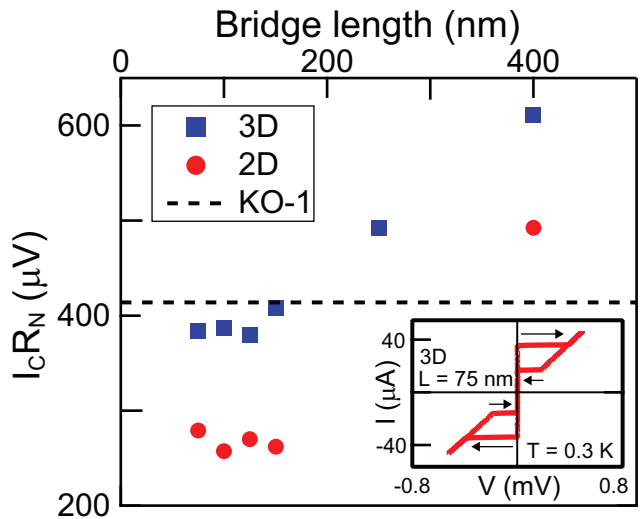} 
\caption{\label{fig2} (Color online) $I_{C}R_{n}$ product as a function of bridge length for both 2D and 3D nanoSQUIDs from several chips. The dashed line in the main figure is the zero temperature prediction for a short metallic weak link connected to ideal phase reservoirs. Inset: a typical I-V curve for a 3D nanoSQUID, consisting of two 75 nm bridges.}
\end{figure}

In the inset of Fig. 2, we present a typical current-voltage (I-V) characteristic of a 3D nanoSQUID with 75 nm long bridges. The I-V characteristic is hysteretic, and the retrapping behavior can be attributed to thermal relaxation \cite{Skocpol}. We observe that the normal state resistance $R_N$ increases linearly with bridge length, with an offset of $\approx 1.5$ $\Omega$ for 3D bridges and $\approx 20 \Omega$ for 2D bridges; these numbers are consistent with the normal state resistances of the banks. We subtract this offset and plot the resulting $I_{C}R_{n}$ product in Fig. 2 for both 2D and 3D nanoSQUIDs. For short 3D devices (L $\leq$ 150 nm), we see that the $I_{C}R_{n}$ product saturates at approximately $380 \mu V$. For comparable 2D devices, the data are clustered around $250 \mu V$. For longer devices, both 2D and 3D, we see a monotonic increase in $I_{C}R_{n}$ with bridge length, as expected for a long superconducting wire \cite{Tinkham}. To interpret the data for short 3D bridges, we can make use of the analytical result from Kulik and Omelyanchuk \cite{Kulik} (KO-1), who calculate the current-phase relation of a weak link with $L\ll \xi$ to be $I(\delta)=(\pi\Delta/e R_{n}) \cos(\delta/2)\tanh^{-1}[\sin(\delta/2)]$ where $\delta$ is the phase difference across the junction and $\Delta$ is the superconducting gap. This formula assumes a diffusive weak link at zero temperature with a phase variation constrained to the bridge region. The dashed line in Fig. 2 indicates the prediction of this relation, $I_C R_N = 1.32 \pi \Delta / 2 e$, which agrees well with our short 3D bridges. For our thin films, we estimate the coherence length in the dirty limit to be $\xi = \sqrt{\hbar D / 2 \pi k_B T_C} \approx 40$ nm, where $D$ is the diffusion constant calculated from the measured film resistivity. Thus, our shortest devices are a few times the coherence length and approximate well a short metallic weak link connected to ideal phase reservoirs.

\begin{figure}[b]
\includegraphics{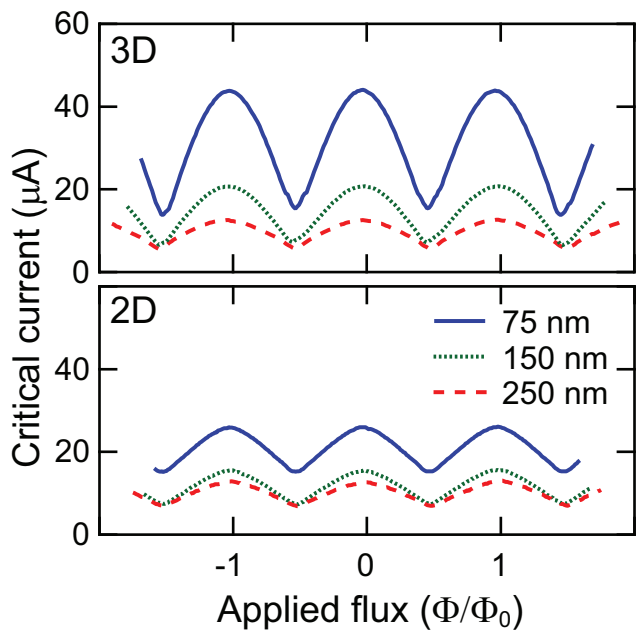}
\caption{\label{fig3} (Color online) Critical current as a function of applied magnetic flux for 75, 150, and 250 nm long nanoSQUIDs. The upper panel are experimental results for nanobridges with 3D banks and the lower panel for 2D banks.}
\end{figure}

To characterize the flux dependence of the nanoSQUIDs we measured I-V curves as a function of applied perpendicular magnetic field. From these data, we extracted the critical current $I_{C}$ as a function of applied flux, and have plotted this quantity for nanoSQUIDS with 75, 150, and 250 nm-long bridges in Fig. 3. We observe that the critical current modulation is deeper and more nonlinear for the 3D devices. The modulation for the 2D devices is quite triangular in shape, indicative of a nearly linear current-phase relation. The data show a modulation pattern which is nearly symmetric about the maximum $I_C$.  Comparing the measured modulation with numerical simulations indicates that the junction critical current asymmetry is typically less than $10\%$. Larger asymmetry would reduce the SQUID modulation depth and thus complicate systematic comparison. The modulation depth in our devices is also not limited by the loop inductance as $\beta_{L} \equiv 2L_{loop}I_{C}/\Phi_{0}\approx 0.1$, where $\Phi_{0}$ is the flux quantum. In Fig. 4, we plot the fractional modulation depth $(I_{C,max}-I_{C,min})/I_{C,max}$ as a function of device length. For short 3D bridges ($L\leq 150$ nm), we observe a modulation depth approaching 70$\%$. In longer bridges, the modulation depth decreases. For very short 2D bridges ($L\leq 100$ nm), we observe a modulation depth below 50$\%$. In intermediate length 2D bridges, the modulation depth increases, and then decreases again for bridges longer than 150 nm.

\begin{figure}[tb]
\includegraphics{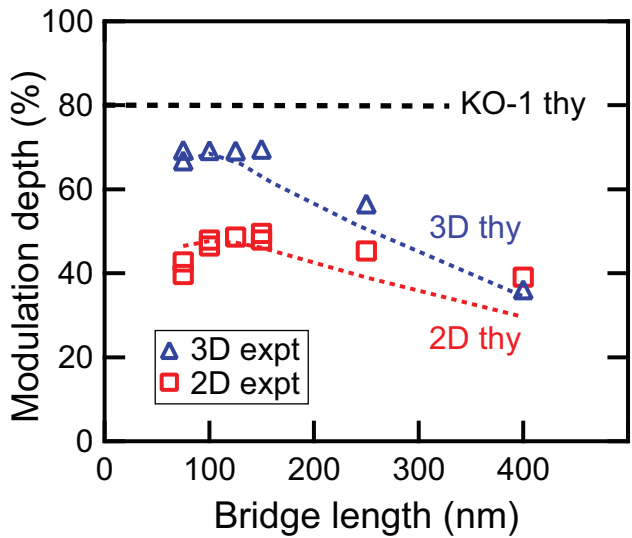}
\caption{\label{fig4}(Color online) Critical current modulation depth as a function of device length for both 3D and 2D nanobridges. The black dashed line is the theoretical prediction for an ideal weak link SQUID at zero temperature with zero loop inductance. The blue and red dashed lines are numerical calculations for our 3D and 2D devices, respectively, including the effect of finite loop inductance.}
\end{figure}

To interpret these results, let us consider a SQUID formed by two symmetric elements with an arbitrary current phase relation $f(\delta)$. In the limit of small loop inductance, the phase difference between the two SQUID junctions $\delta_{2}-\delta_{1}$ equals the applied flux $2\pi \Phi_{m}/\Phi_{0}$.  The current $I$ through the SQUID is the sum of the currents in each arm $I=f(\delta_{1})+f(\delta_{2})=f(\delta_{1})+f(\delta_1 + 2\pi \Phi_{m}/\Phi_{0})$. For a tunnel junction based SQUID, $f(\delta)=I_{0} \sin(\delta)$; maximizing the current equation above gives the SQUID critical current $I_C=2I_{0}| \cos(\pi\Phi_{m}/\Phi_{0})|$ where $I_{0}$ is the critical current of each junction. The critical current modulates from 0 to $2I_{0}$ and hence the modulation depth is 100$\%$. For a linear current phase relation, expected for a long wire, the maximum modulation depth is 25$\%$. In the case of the KO-1 relation cited earlier, the maximum modulation depth is 81$\%$. Thus, our short 3D bridges approach the KO-1 result, while the longest devices tend towards 25$\%$ modulation as the CPR becomes increasingly linear. Also plotted in Fig. 4 are the modulation depths predicted from our numerically computed CPRs \cite{vijay_wire}. These predictions take into account loop inductance, using the condition $2 \pi n = \delta_1 - \delta_2 +  2\pi \Phi_{m}/\Phi_{0} + \pi L_{loop} [ f(\delta_1) - f(\delta_2)] / \Phi_0 $. The quantitative agreement with short 2D and 3D devices is excellent, except for the shortest 2D devices, which modulate less than predicted. We attribute this to the fact the phase gradient in short 2D bridges leaks into the banks of the device.  Thus, for two weak links in close proximity it is difficult to impose a phase difference between the two arms of the SQUID \cite{vijay_wire}, and modulation is suppressed. Our calculations are for isolated junctions and a numerical calculation of the full two junction SQUID is required to reproduce this effect \cite{hasselbach}. 

In summary, using low-inductance symmetric dc SQUIDs, we have demonstrated that nanobridge Josephson junctions contacted with 3D banks exhibit greater nonlinearity than those with 2D banks.  Transport in 3D bridges up to several coherence lengths long is well described by the KO-1 formula for an ideal, short, metallic weak link. Our experimental results give good quantitative agreement with a numerical solution of the Usadel equations for the precise geometry under study. As such, 3D nanobridge devices should have sufficient nonlinearity for use in high gain amplifier and qubit circuits traditionally employing tunnel junctions.  These devices could also be used for dispersive, high-speed magnetometry by embedding them in a microwave resonator\cite{hatridge}. For magnetometry, 3D nanoSQUIDs should have higher flux sensitivity than 2D devices, but perhaps with a reduced tolerance to in-plane magnetic fields. Further study is needed to determine the variation of nonlinearity and field tolerance with bank thickness.  

The authors thank Gary Williams for assistance with device imaging.  Financial support was provided by AFOSR under Grant FA9550-08-1-0104 (RV) and DARPA under Grant N66001-09-1-2112 (IS). EMLF gratefully acknowledges an NDSEG fellowship awarded by DoD, Air Force Office of Scientific Research, 32 CFR 168a. DHS gratefully acknowledges support from a Hertz Foundation Fellowship endowed by Big George Ventures.


\begin{thebibliography}{12}
\expandafter\ifx\csname natexlab\endcsname\relax\def\natexlab#1{#1}\fi
\expandafter\ifx\csname bibnamefont\endcsname\relax
  \def\bibnamefont#1{#1}\fi
\expandafter\ifx\csname bibfnamefont\endcsname\relax
  \def\bibfnamefont#1{#1}\fi
\expandafter\ifx\csname citenamefont\endcsname\relax
  \def\citenamefont#1{#1}\fi
\expandafter\ifx\csname url\endcsname\relax
  \def\url#1{\texttt{#1}}\fi
\expandafter\ifx\csname urlprefix\endcsname\relax\def\urlprefix{URL }\fi
\providecommand{\bibinfo}[2]{#2}
\providecommand{\eprint}[2][]{\url{#2}}

\bibitem[{\citenamefont{Lam and Tilbrook}(2003)}]{CSIRO}
\bibinfo{author}{\bibfnamefont{S.~K.~H.}} \bibnamefont{Lam} \bibnamefont{and}
\bibinfo{author}{\bibfnamefont{D.~L.}} \bibnamefont{Tilbrook},
	\bibinfo{journal} {Appl. Phys. Lett.} \textbf{\bibinfo{volume}{82}},
	\bibinfo{pages}{1078} (\bibinfo{year}{2003}).
	
\bibitem[{\citenamefont{Troeman et~al.}(2007)}]{Hilgenkamp}
\bibinfo{author}{\bibfnamefont{A.~G.~P.}} \bibnamefont{Troeman},
\bibinfo{author}{\bibfnamefont{H.}} \bibnamefont{Derking},
\bibinfo{author}{\bibfnamefont{B.}} \bibnamefont{Borger},
\bibinfo{author}{\bibfnamefont{J.}} \bibnamefont{Pleikies},
\bibinfo{author}{\bibfnamefont{D.}} \bibnamefont{Veldhuis}, \bibnamefont{and}
\bibinfo{author}{\bibfnamefont{H.}} \bibnamefont{Hilgenkamp},
	\bibinfo{journal} {Nano Lett.} \textbf{\bibinfo{volume}{7}},
	\bibinfo{pages}{2152} (\bibinfo{year}{2007}).	

\bibitem[{\citenamefont{Wernsdorfer}(2009)}]{Wernsdorfer_review}
\bibinfo{author}{\bibfnamefont{W.}} \bibnamefont{Wernsdorfer},
	\bibinfo{journal} {Sup. Sci. Tech.} \textbf{\bibinfo{volume}{22}},
	\bibinfo{pages}{064013} (\bibinfo{year}{2009}).

\bibitem[{\citenamefont{Simmonds et~al.}(2004)}]{Martinis}
\bibinfo{author}{\bibfnamefont{R.~W.}} \bibnamefont{Simmonds},
\bibinfo{author}{\bibfnamefont{K.~M.}} \bibnamefont{Lang},
\bibinfo{author}{\bibfnamefont{D.~A.}} \bibnamefont{Hite},
\bibinfo{author}{\bibfnamefont{S.}} \bibnamefont{Nam},
\bibinfo{author}{\bibfnamefont{D.~P.}} \bibnamefont{Pappas}, \bibnamefont{and}
\bibinfo{author}{\bibfnamefont{J.~M.}} \bibnamefont{Martinis},
	\bibinfo{journal} {Phys. Rev. Lett.} \textbf{\bibinfo{volume}{93}},
	\bibinfo{pages}{077003} (\bibinfo{year}{2004}).

\bibitem[{\citenamefont{Likharev}(1979)}]{Likharev}
\bibinfo{author}{\bibfnamefont{K.~K.}} \bibnamefont{Likharev},
	\bibinfo{journal} {Rev. Mod. Phys.} \textbf{\bibinfo{volume}{51}},
	\bibinfo{pages}{101} (\bibinfo{year}{1979}).

\bibitem[{\citenamefont{Vijay et~al.}(2009)}]{vijay_wire}
\bibinfo{author}{\bibfnamefont{R.}} \bibnamefont{Vijay},
	\bibinfo{author}{\bibfnamefont{J.~D.}} \bibnamefont{Sau},
	\bibinfo{author}{\bibfnamefont{Marvin~L.}} \bibnamefont{Cohen}, \bibnamefont{and}
	\bibinfo{author}{\bibfnamefont{I.}} \bibnamefont{Siddiqi},
	\bibinfo{journal} {Phys. Rev. Lett.} 
	\textbf{\bibinfo{volume}{103}}, \bibinfo{pages}{087003} (\bibinfo{year}{2009}).
	
\bibitem[{\citenamefont{Skocpol et~al.}(1974)}]{Skocpol}
\bibinfo{author}{\bibfnamefont{W.~J.} \bibnamefont{Skocpol}}, 
  \bibinfo{author}{\bibfnamefont{M.~R.} \bibnamefont{Beasley}}, \bibnamefont{and}
  \bibinfo{author}{\bibfnamefont{M.} \bibnamefont{Tinkham}},
  \bibinfo{journal}{J. Appl. Phys.}
  \textbf{\bibinfo{volume}{45}}, \bibinfo{pages}{4054} (\bibinfo{year}{1974}).

	
\bibitem[{\citenamefont{Kulik and Omel'Yanchuk}(1975)}]{Kulik}
\bibinfo{author}{\bibfnamefont{I.~O.} \bibnamefont{Kulik}} \bibnamefont{and}
  \bibinfo{author}{\bibfnamefont{A.~N.} \bibnamefont{Omel'Yanchuk}},
  \bibinfo{journal}{JETP Lett.}
  \textbf{\bibinfo{volume}{21}}, \bibinfo{pages}{96} (\bibinfo{year}{1975}).
  
\bibitem[{\citenamefont{Tinkham}(1996)}]{Tinkham}
\bibinfo{author}{\bibfnamefont{M.}} \bibnamefont{Tinkham},
	\emph{\bibinfo{title} {Introduction to Superconductivity}},
	(\bibinfo{publisher}{McGraw-Hill}, \bibinfo{year}{1996}).
  
\bibitem[{\citenamefont{Hasselbach et~al.}(2002)\citenamefont{Hasselbach, Mailly, and Kirtley}}]{hasselbach}
\bibinfo{author}{\bibfnamefont{K.} \bibnamefont{Hasselbach}},
  \bibinfo{author}{\bibfnamefont{D.} \bibnamefont{Mailly}}, \bibnamefont{and}
  \bibinfo{author}{\bibfnamefont{J.~R.} \bibnamefont{Kirtley}},
  \bibinfo{journal}{J. Appl. Phys.} \textbf{\bibinfo{volume}{91}},
  \bibinfo{pages}{4432} (\bibinfo{year}{2002}).
  
\bibitem[{\citenamefont{Hatridge et~al.}(2010)}]{hatridge}
\bibinfo{author}{\bibfnamefont{M.~J.} \bibnamefont{Hatridge}},
  \bibinfo{author}{\bibfnamefont{R.} \bibnamefont{Vijay}},
  \bibinfo{author}{\bibfnamefont{D.~H.} \bibnamefont{Slichter}},
  \bibinfo{author}{\bibfnamefont{John} \bibnamefont{Clarke}}, \bibnamefont{and}
  \bibinfo{author}{\bibfnamefont{I.} \bibnamefont{Siddiqi}},
  \bibinfo{journal}{arXiv:1003.2466v1}.
  
\end{thebibliography}
\end{document}